\documentclass[aip,jcp,reprint,amsmath,superscriptaddress]{revtex4-1}
\usepackage{amsmath}
\usepackage[latin1]{inputenc}
\usepackage{graphicx}
\usepackage[]{hyperref}

\usepackage{epsfig}

\usepackage{color}
\usepackage{xcolor}
\hypersetup{
    colorlinks,
    linkcolor={red!50!black},
    citecolor={blue!50!black},
    urlcolor={blue!80!black}
}

\begin{document}
\title{Percolation in binary and ternary mixtures of patchy colloids}
\begin{abstract}
We investigate percolation in binary and ternary mixtures of patchy colloidal particles theoretically and using Monte Carlo simulations. Each particle has three identical patches, with distinct species having different types of patch. Theoretically we assume tree-like clusters and calculate the bonding probabilities using Wertheim's first-order perturbation theory for association.
For ternary mixtures we find up to eight fundamentally different percolated states. The states differ in terms of the species and pairs 
of species that have percolated. The strongest gel is a trigel or tricontinuous gel, in which each of the three species has percolated. 
The weakest gel is a mixed gel in which all of the particles have percolated, but none of the species percolates by itself. The competition between entropy of mixing and internal energy of 
bonding determines the stability of each state. Theoretical and simulation results are in very good agreement. The only significant difference is the temperature at the percolation threshold, which is overestimated by the theory due to the absence of closed loops in the theoretical description.
\end{abstract}

\author{Felix Seiferling}
\affiliation{Theoretische Physik II, Physikalisches Institut,
  Universit{\"a}t Bayreuth, D-95440 Bayreuth, Germany}

\author{Daniel de las Heras}
\email{delasheras.daniel@gmail.com}
\affiliation{Theoretische Physik II, Physikalisches Institut,
  Universit{\"a}t Bayreuth, D-95440 Bayreuth, Germany}

\author{Margarida M. Telo da Gama}
\affiliation{Departamento de F\'{\i}sica e Centro de F\'{\i}sica Te\'orica e Computacional, Faculdade de Ci\^encias, Universidade de Lisboa, Campo Grande, P-1749-016, Lisbon, Portugal}

\maketitle

\section{Introduction}

The interaction between patchy colloids is valence limited, specific, and directional due to the presence of patches or interaction sites on the surface of the colloids.
Patchy colloids are excellent candidates to design new materials with properties on-demand by controlling the microscopic details of the interparticle interaction.
The distribution,
number, and types of patches can be tuned~\cite{glotzer2007anisotropy,in2,0953-8984-25-19-193101,doi:10.1021/acsnano.5b07901} such that patchy colloids self-assemble into new structures that are not found in isotropically interacting colloids. Examples
are the directed self-assembly of patchy colloids into a kagome lattice~\cite{chen2011directed}, polyhedra~\cite{polyh}, and micelles~\cite{kraft2012surface}. 

In addition to complex regular structures, patchy colloids also form stable gels. The number of patches is a key ingredient determining the location of the liquid-gas critical
point~\cite{:/content/aip/journal/jcp/128/14/10.1063/1.2888997}. In systems of patchy colloids with low valence the liquid-gas phase separation region shrinks drastically,
and it is possible to find stable liquids at very low densities and temperatures~\cite{emptyscio,herasNF}.
These, equilibrium, low density states are gels in which the patchy colloids assemble into amorphous percolated networks.

Mixtures of patchy particles provide an additional handle for tuning the morphology of the gel. In~\cite{herasBC} we predicted theoretically the 
existence of new gel structures in binary mixtures with intriguing percolation properties, such a bicontinuous gel or
bigel. Here, two interpenetrating networks, each made of one type of colloids, span the system volume. Other gel structures are possible, such as mixed gels where the mixture is percolated but neither species percolates independently~\cite{herasBC}.

Bigel-like percolated states are not specific of patchy colloidal systems. Bicontinuous structures have been reported in e.g., polymer blends~\cite{coco1,coco2},
mixtures of dipolar colloidal particles~\cite{Goyal1,Goyal2}, fumed silica-based systems~\cite{C4RA15437A}, mixtures of an aqueous gel and an oleogel~\cite{pharma,pharma2}, DNA coated colloids~\cite{Varrato20112012,C3SM52558A}, and mixtures of proteins~\cite{C5TB00131E}.

In this work, we investigate the structure of equilibrium gels in binary and ternary mixtures of patchy colloids. We test the validity of our
extension of the Flory-Stockmayer theory of polymerization~\cite{flory,stock} to multicomponent
systems with different types of patches~\cite{herasNF} by comparing theoretical with Monte Carlo simulation results. The latter agree semi-quantitatively with the theoretical predictions 
and corroborate our previous findings~\cite{herasBC} of bigels and mixed gels in binary systems of patchy colloids. 
Additionally, we extend the analysis of the percolation of patchy colloids to ternary mixtures both theoretically and using computer simulations. We find eight fundamentally different types
of percolated states
that we classify according to their strength against breaking bonds. The strongest gel is a trigel or tricontinuous gel in which there are
three interpenetrated networks, each one made of one type of colloids. The rich phenomenology of the ternary system 
can be understood as a competition between the entropy of mixing and the internal energy of bonding. At high temperature the entropy of mixing, which
favors interspecies bonds, dominates and promotes the formation of different types of mixed gels. At low temperatures, however, the internal energy of bonding
is the most important contribution to the free energy. Here, we observe the occurrence of strong gels such as bigels and trigels 
provided that the intraspecies bonds are stronger than the interspecies ones. The stability of each type of gel can be controlled by changing the
temperature and composition of the mixture.

\section{Model and methods}
\begin{figure}
\includegraphics[width=0.9\columnwidth]{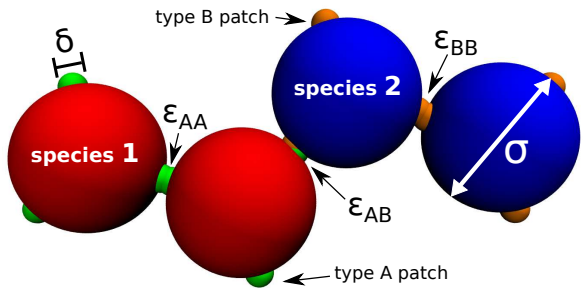}
\caption{Illustration of the model. Patchy colloids are modelled as hard spheres of diameter $\sigma$ with three bonding sites
of size $\delta$ on the surface. The only difference between species is the type of their bonding sites: type A for species $1$, type B for species
$2$, and type C for species $3$ (not shown). A bond between patches of type $\alpha$ and $\beta$ decreases the energy of the system by a
quantity $\epsilon_{\alpha\beta}$. }
\label{fig1}
\end{figure}

The model patchy colloidal particles, see Fig.~\ref{fig1}, consists of a hard core (hard sphere of diameter $\sigma$) with three interaction sites or
patches on the surface. The patches are spheres of size $\delta$ centered at the surface of the hard core, 
uniformly distributed along the equator of the particles, i.e. 
the angle between two patches is $120^\circ$. We set $\delta=0.5(\sqrt{5-2\sqrt{3}}-1)\sigma\approx0.12\sigma$, which is the largest
value that guarantees single bonding of each patch. In addition, the geometry of the particles ensures that two colloids
cannot form more than one bond between them. The interaction
between patches $\phi_{\alpha\beta}$ is a square well potential, i.e. if two patches of type $\alpha$ and $\beta$ overlap, the energy
of the system decreases by a quantity $\epsilon_{\alpha\beta}$:
\begin{equation}
\phi_{\alpha\beta}(1,2)=
\begin{cases} 
-\epsilon_{\alpha\beta}, \text{patches overlap}  \\ 
0, \text{ patches do not overlap},
\end{cases}
\label{potential}
\end{equation}
where $1,2$ indicates the position and orientations of colloids $1$ and $2$.

We consider binary and ternary mixtures. The only difference
between the species is the type of their patches. In the case of binary mixtures the particles of species $1$ $(2)$ have patches
of type A (B). For ternary mixtures we add a third species with patches of type C.
\subsection{Theory}
We use a generalization of the Flory-Stockmayer theory of polymerization~\cite{flory,stock} to study the percolation transition. 
The theory neglects bond correlations, and therefore assumes that all clusters are tree-like.
For details of the theory for multicomponent systems with different types of patches we refer the reader to Refs.~\cite{herasNF,herasBC}. 
Here, we describe the theory for our patchy particle model: each particle has three identical patches, and there are three types of particles.
The binary system may be obtained as a particular case.

\begin{figure}
\includegraphics[width=0.9\columnwidth]{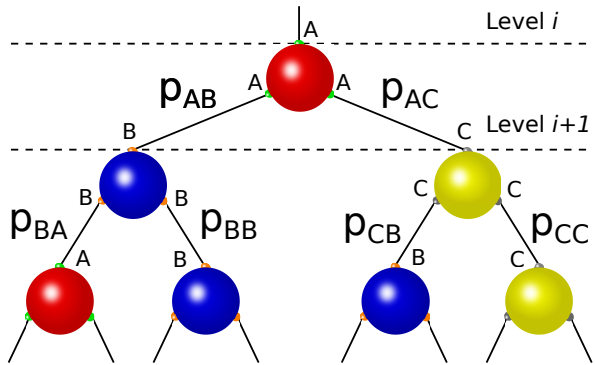}
\caption{Schematic tree-like cluster in a ternary mixture. Each particle of species 1 (red), 2 (blue), and 3 (yellow) has 
three patches of type A, B, and C, respectively. $p_{\alpha\beta}$ is the probability of bonding a site $\alpha$ to a site $\beta$,
with $\alpha,\beta=\{\text{A,B,C}\}$.}
\label{fig2}
\end{figure}

A schematic tree-like cluster is shown in Fig.~\ref{fig2}. Let $n_{i+1,\alpha}$ be the number of patches of type $\alpha$ that are bonded
at level $i+1$ of the tree-like cluster, see  Fig.~\ref{fig2}. $n_{i+1,\alpha}$ is related to the number of bonds that are bonded at the
previous level $i$ via
\begin{eqnarray}
&&n_{i+1,\text A}=n_{i,\text A}2p_{\text{AA}}+n_{i,\text B}2p_{\text{BA}}+n_{i,\text C}2p_{\text{CA}},\nonumber\\
&&n_{i+1,\text B}=n_{i,\text A}2p_{\text{AB}}+n_{i,\text B}2p_{\text{BB}}+n_{i,\text C}2p_{\text{CB}},\nonumber\\
&&n_{i+1,\text C}=n_{i,\text A}2p_{\text{AC}}+n_{i,\text B}2p_{\text{BC}}+n_{i,\text C}2p_{\text{CC}},
\label{prop}
\end{eqnarray}
where $p_{\alpha\beta}$ is the probability of bonding a site $\alpha$ to a site $\beta$. Note that, in general, $p_{\alpha\beta}\neq p_{\beta\alpha}$
since $p_{\alpha\beta}$ is proportional to the number of available patches of type $\beta$ and hence to the number of particles with $\beta$ patches, as we will see.
The factor of two in all the terms
of the above equation reflects the fact that one patch is bonded at level $i$ by construction, and therefore only two patches are available to form bonds
at the next level $i+1$. We can rewrite Eq.~\eqref{prop} in matrix form
\begin{equation}
\mathbf{n_{i+1}}=\tilde{\mathbf T}\mathbf{n_i}=\tilde{\mathbf T}^i\mathbf{n_0},
\end{equation}
where the components of the vector $\mathbf{n_{i}}$ are $n_{i,\alpha}$, and the percolation matrix $\tilde{\mathbf T}$ is
\begin{eqnarray}
\tilde{\mathbf T}=2\left( \begin{array}{ccc}
p_{\text{AA}} & p_{\text{BA}} & p_{\text{CA}} \\
p_{\text{AB}} & p_{\text{BB}} & p_{\text{CB}} \\
p_{\text{AC}} & p_{\text{BC}} & p_{\text{CC}} \end{array} \right).
\end{eqnarray}
The system percolates if the number of bonds diverges in the limit $i\rightarrow\infty$, or equivalently if $|\lambda|>1$ with $\lambda$ the largest eigenvalue of the percolation 
matrix, see \cite{PhysRevE.81.010501} for more details. In order to establish if subsets of the
whole system are also percolated we have to consider a percolation matrix that
includes only the desired bonds. For example, the pair formed by species $1$ (patches of type A) and species 2 (patches of type B)
has percolated if the largest eigenvalue of the matrix:
\begin{eqnarray}
\tilde{\mathbf T}'=2\left( \begin{array}{cc}
p_{\text{AA}} & p_{\text{BA}} \\
p_{\text{AB}} & p_{\text{BB}} \end{array} \right).
\end{eqnarray}
is larger than one. The same procedure applied to a single species leads to the known percolation threshold
for tree-like clusters of identical particles with only one type of patch:  $p_{\alpha\alpha}\ge0.5=1/(f-1)$, with $f$ the number of patches.

We obtain the probabilities $p_{\alpha\beta}$ from Wertheim's first order perturbation theory~\cite{wertheim}, which also neglects correlations between bonds.
The probability of bonding for a site $\alpha$ is
\begin{equation}
\pi_\alpha=1-X_\alpha=\sum_{\gamma={\text{A,B,C}}}p_{\alpha\gamma},
\label{Pa}
\end{equation}
where $X_\alpha$ is the probability that a site $\alpha$ is not bonded. The probabilities $X_\alpha$ are obtained from Wertheim's theory. For our model, we find 
(see \cite{herasBC} for details)
\begin{equation}
X_\alpha=\left[1+3\eta\sum_{i=1}^{3}x^{(i)}X_{\beta}\Delta_{\alpha\beta}\right]^{-1},
\label{Xa}
\end{equation}
where $x^{(i)}$ is the composition of species $i$, the subscript $\beta$ denotes the type of patches of species $i$ (i.e., $\beta=\text A$ if $i=1$, $\beta=\text B$ if $i=2$, and 
$\beta=\text C$ if $i=3$), $\eta$ is the
total packing fraction of the system, and 
\begin{equation}
\Delta_{\alpha\beta}=\Delta_{\beta\alpha}=\frac{1}{v_{\text{s}}}\int g_{\text{HS}}(r)[\exp{(-\phi_{\alpha\beta}(1,2)/k_{\text{B}}T)}-1]d(12).
\end{equation}
Here, $v_{\text{s}}$ is the volume of the hard-core of the patchy colloids, $k_{\text{B}}$ is the Boltzmann constant, and $g_{\text{HS}}(r)$ is the
radial distribution function for hard-spheres at a distance $r$. The integral is calculated over all possible orientations and separations of the colloids.
Finally, we approximate $g_{\text{HS}}$ by its contact value $A_0(\eta)$ (the patches are very small compared to the size of the colloids). Hence,
\begin{equation}
\Delta_{\alpha\beta}\approx\frac{v_{\text b}}{v_{\text s}}A_0(\eta)[\exp{(\epsilon_{\alpha\beta}/k_{\text{B}}T)}-1],
\end{equation}
with
\begin{equation}
A_0(\eta)=\frac{1-\eta/2}{(1-\eta)^3},
\end{equation}
and $v_b$ is the bonding volume, i.e., the volume in which two colloids can form a bond averaged over all possible orientations of the colloids. 
For our model of the patches~\cite{mcwt} $v_{\text{b}}\approx3.32\cdot10^{-4}\sigma^3$.

Analyzing term-by-term Eqs.~\eqref{Pa} and~\eqref{Xa} we obtain
\begin{equation}
p_{\alpha\beta}=x^{(i)}3X_\alpha X_\beta\eta\Delta_{\alpha\beta},
\end{equation}
which has an intuitive explanation. The probability of bonding a site $\alpha$ to a site $\beta$, $p_{\alpha\beta}$, is proportional to: (i) the density of available patches of type $\beta$ which is  $3x^{(i)}X_\beta\eta/v_{\text s}$, and (ii) the probability that a patch $\alpha$ is available (not bonded) which is $X_\alpha$. The parameter $\Delta_{\alpha\beta}$ controls the strength of the bond.

We have also obtained the probabilities directly from Monte Carlo simulations and used them to compute the percolation matrix. We compare both methods in the Results section.

\subsection{Simulation}
We use Monte Carlo simulations in the canonical ensemble, that is, we fix the total number of particles of each species $N_i$, the temperature $T$,
and the system volume $V$. The simulation box is a cube with periodic boundary conditions. The total number of particles in the simulation box is
$N=300$ and the packing fraction $\eta=0.30$. We have also computed selected cases with $N=600$ and the same packing fraction and found no significant differences. 
Therefore, we expect that finite size effects are irrelevant.

In each simulation we collect data from $10^7$ Monte Carlo steps (MCS). Each MCS is an attempt to move and rotate all particles in the system. In addition to translations
and rotations of the particles we introduced a particle swap move~\cite{PhysRevE.63.045102}. The swap move randomly selects two particles 
and interchanges their coordinates (spatial and orientational). The swap move is accepted according to the standard Metropolis criterion and satisfies detailed balance. 
We perform $0.1N$ swap attempts after every MCS.
The particle swap is very useful to prevent trapping of the particles in local minima of the energy landscape, especially at low temperatures. Before collecting the data,
we equilibrate the system by running MCS until the energy fluctuates around a minimum value. Typically $\sim 10^6-10^7$ MCS are used for equilibration, depending on
the temperature.

For each composition we start by running a simulation at high temperature. Next we decrease $T$ using the last configuration of the previous simulation as the initial state.
We repeat this process for several $T$s. In order to check we have reached the equilibrium states we carried out the inverse process for selected cases. That is, we started at low
$T$ and increased the temperature. We have found the same states both by increasing or decreasing $T$.

The main output of the simulation is the cluster size distribution,
\begin{equation}
s(n)=N_{\text c}(n)/N,
\end{equation}
where $N_{\text c}(n)$ is the number of clusters of size $n$. We compute $s(n)$ for the whole mixture, for each species $s^{(i)}(n)=N_{\text c}^{(i)}/N_i$, and in the case
of ternary mixtures for pairs of species $s^{(i+j)}(n)$. Using the cluster size distribution we can compute the probability of finding a particle
in a cluster of size $n$:
\begin{equation}
P_n=ns(n),
\end{equation}
and the moments of $s(n)$. We use the second moment of the cluster size distribution
\begin{equation}
N_{\text w}=\frac{1}N\sum_nn^2s(n),
\end{equation}
to characterize the percolation transition.  We set the percolation threshold at $0.5$, i.e., we consider the whole system
has percolated if $N_{\text w}>0.5$. In order to characterize the type of percolated state we also analyze the second moments of
the cluster size distributions of single species and pairs of species. E.g., the species $i$ has percolated if $N^{(i)}_{\text w}=\frac{1}{N_i}\sum_nn^2s^{(i)}(n)>0.5$.

\section{Results}
\subsection{Binary mixtures}
\begin{figure}
\includegraphics[width=0.8\columnwidth]{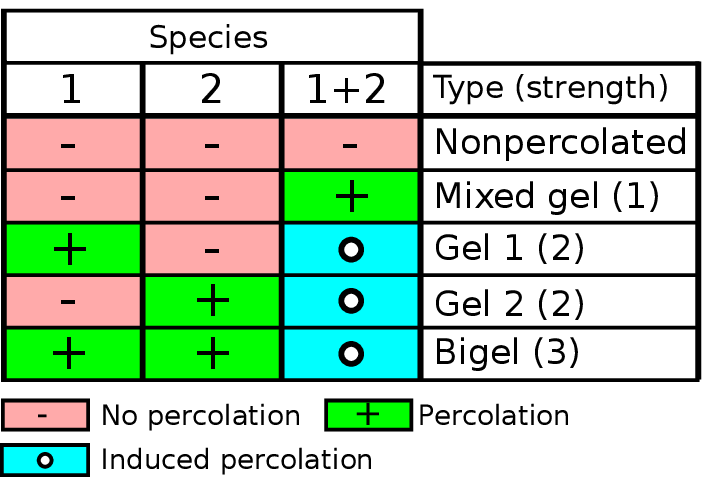}
\caption{Types of gels of a binary mixture of patchy particles. The number between brackets after
the name of the gel indicates its strength against breaking bonds (the higher this number the stronger the gel).}
\label{fig3}
\end{figure}

We start by describing the results for a binary mixture, a system that we have previously studied using Wertheim's theory~\cite{herasBC}. All possible gels
are listed in Fig.~\ref{fig3}. We order the gels according to their relative strength against breaking bonds. The weakest gel is a mixed gel (MG). In a MG
the mixture has percolated, but neither species $1$ nor species $2$ have percolated. The removal of one of the species will break the connectivity of the system. 
Next we have a standard gel. Here, one species percolates, and as a consequence the mixture also percolates (in the mixture we add new
bonds to a system that has percolated). The percolation of the mixture is driven (induced) by the percolation of one of the species. 
A standard gel is stronger than a mixed gel in the sense that we can remove one species, the one that has not percolated,
and the system remains in a percolated state. The strongest gel is a bicontinuous (BG) gel, where both species percolate independently. We can remove one species, either
$1$ or $2$, and the other species remains in a percolated state.

\begin{figure*}
\includegraphics[width=0.95\textwidth]{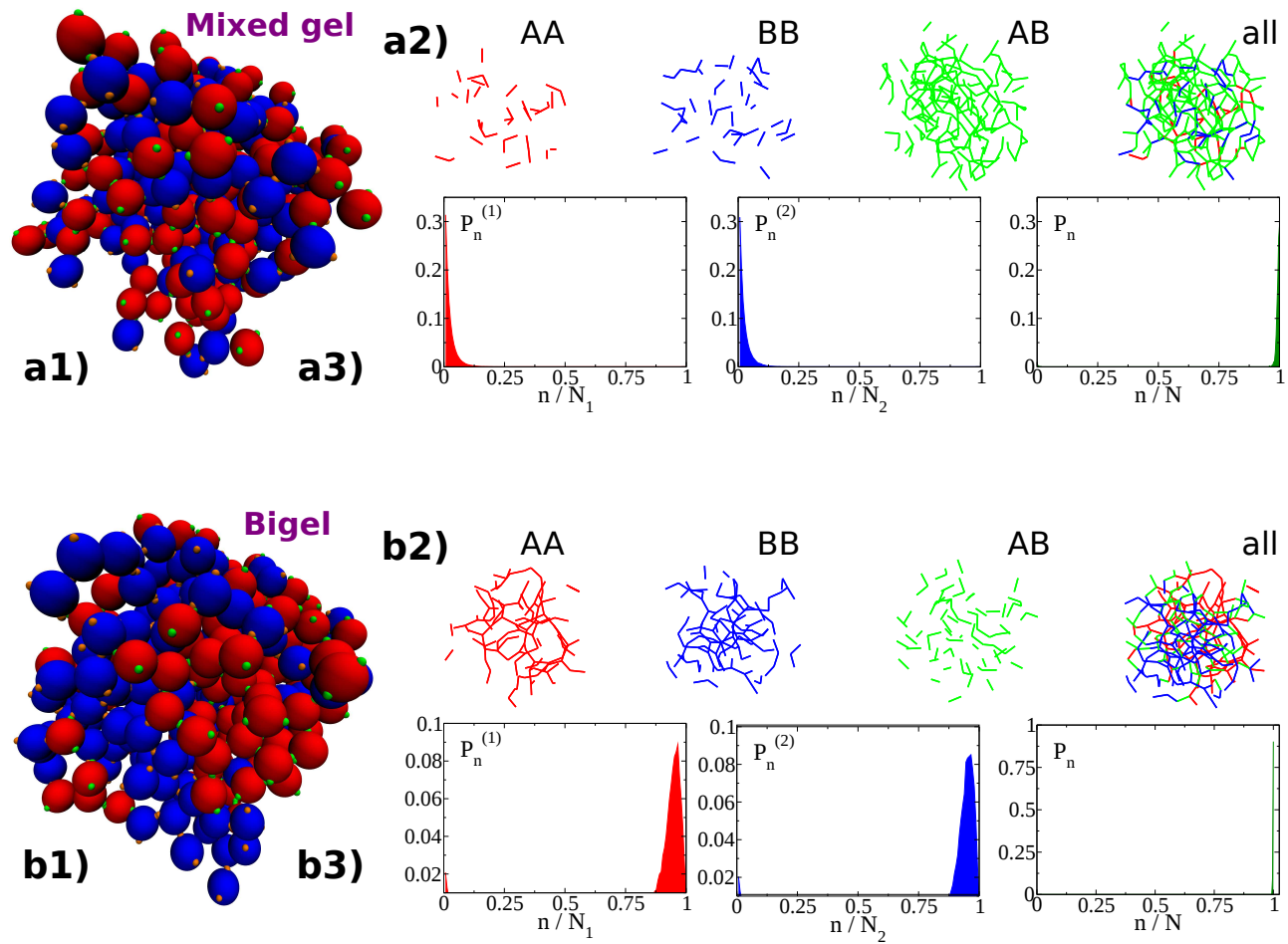}
\caption{Simulation results for a mixed gel (upper panels) at
$\eta=0.3$, $x=0.5$, and $k_BT/\epsilon_{AA}=0.10$. The bonding energies
are $\epsilon_{BB}=\epsilon_{AA}$ and $\epsilon_{AB}=1.05\epsilon_{AA}$. (a1) 
Typical simulation snapshot. (a2) Graphical representation of the bonds; each
line connects the center of mass of two particles bonded through AA bonds (left most), BB bonds
(second left), AB bonds (second right), and through any type of bonds (right most). (a3) Probability of
finding a particle in a cluster of size $n$, $P_n=ns(n)$, for particles of species $1$ (left), $2$ (middle), and for the whole mixture (right) as 
a function of the cluster size (normalized with the corresponding number of particles). The bottom panels depict similar results for a bicontinuous gel at $\eta=0.30$, $x=0.5$, $k_BT/\epsilon_{AA}=0.08$
and bonding energies $\epsilon_{BB}=\epsilon_{AA}$, $\epsilon_{AB}=0.90\epsilon_{AA}$.
}
\label{fig4}
\end{figure*}

The relative stability of the different gels depends on many factors, such as temperature,
composition, packing fraction and ratio between the different bonding energies. In Fig.~\ref{fig4}
we show an example of a mixed and a bicontinuous gel. Both mixtures are symmetric ($\epsilon_{AA}=\epsilon_{BB}=1$), and
the packing fraction ($\eta=0.3$), and composition ($x^{(i)}=0.5$) are also the same. A mixed gel (see a snapshot in Fig.~\ref{fig4}a1) can be easily stabilized by making the interspecies
bonds stronger than the intraspecies bonds.  In this way both the entropy of mixing and the internal energy of bonding favour the formation of interspecies bonds, which are
dominant in a mixed gel (see Fig.~\ref{fig4}a2). The probabilities of finding a particle in a cluster of $n$ identical particles $P_n^{(i)}$
(Fig.~\ref{fig4}a3) show that only small clusters of identical particles are present, whereas the probability of finding a particle in a cluster of $n$ particles (independently of the particle type)
clearly indicates that the mixture has percolated. If interspecies bonds are weaker than intraspecies bonds, i.e. $\epsilon_{AB}<\epsilon_{\alpha\alpha}$ with $\alpha={A,B}$,
then the energy of bonding may overcome the entropy of mixing. In this case a bicontinuous gel appears (see a snapshot in Fig.~\ref{fig4}b1). Here, intraspecies bonds dominate (see Fig.~\ref{fig4}b2),
and the cluster size distribution indicates that both species independently and the mixture have percolated (see Fig.~\ref{fig4}b3). 

\begin{figure}
\includegraphics[width=0.85\columnwidth]{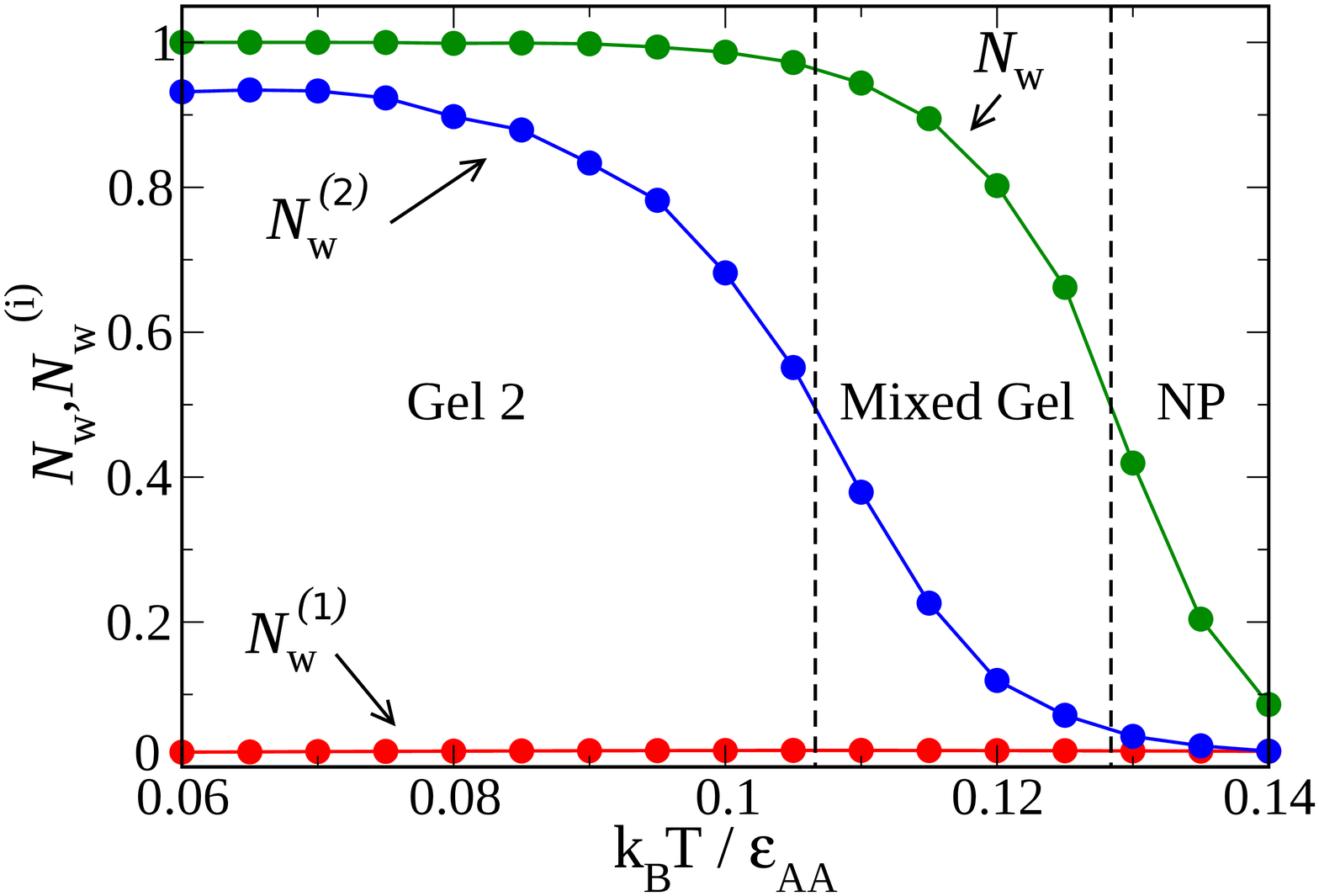}
\caption{Second moment of the cluster size distribution for each species $N_{\text w}^{(i)}$ and for the mixture $N_{\text w}$ as a function of the scaled temperature $k_{\text B}T/\epsilon_{\text{AA}}$.
Symmetric mixture ($\epsilon_{AA}=\epsilon_{BB}$) with ${\epsilon_{AB}}=1.05\epsilon_{AA}$, composition $x=0.2$, and packing fraction $\eta=0.30$.}
\label{fig5}
\end{figure}

The ratio between inter- and intra-species bonding energies plays a major role in determining the state of the sample but it is not the only relevant variable. Temperature, packing fraction,
and composition are also important. We illustrate this in Fig.~\ref{fig5}, where the second moment of the cluster size distributions as a function of the temperature is plotted for a symmetric
mixture with $\epsilon_{\text{AB}}=1.05\epsilon_{\text{AA}}$ and $x=N_1/N=0.2$. At high temperatures there are only a few bonds and the system has not percolated. Percolation occurs
at $k_{\text B}T_{\text p}/\epsilon_{\text{AA}}\approx0.13$. The system is always percolated below $T_{\text p}$ but the type of percolation changes. In a range of temperatures below $T_{\text p}$ there is a
mixed gel; only the two species together percolate. However, at low temperatures the state changes to a gel of species $2$. That is, the particles of species $2$, which are the majority,
form a percolated network. The whole mixture also percolates as addition of another species only adds new bonds to the network. 

\begin{figure}
\includegraphics[width=0.9\columnwidth]{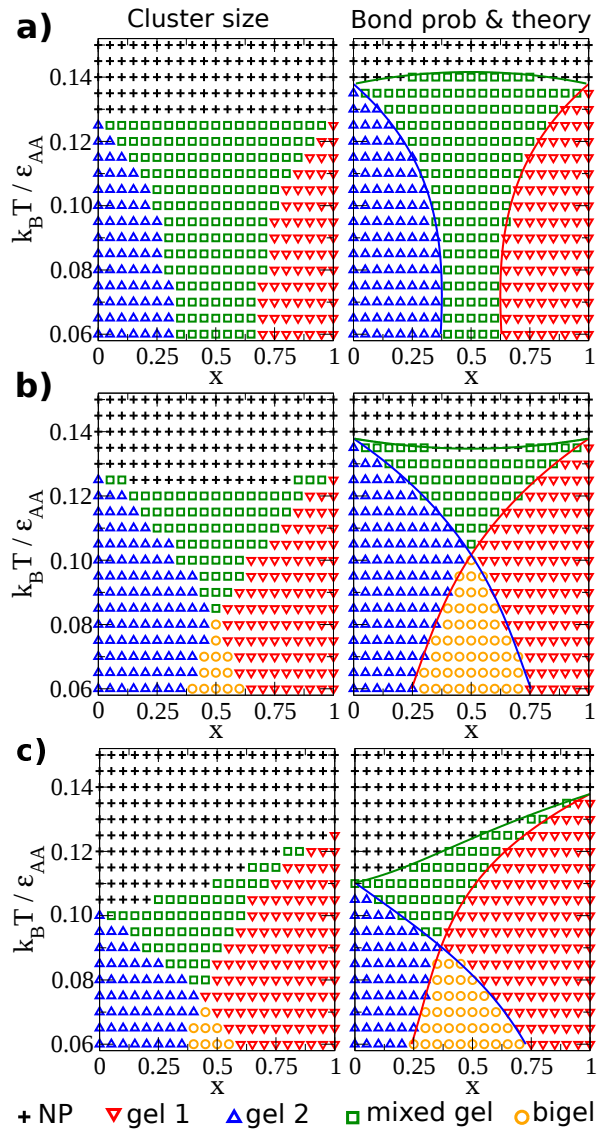}
\caption{Percolation diagrams in the plane of composition $x$ and scaled temperature $k_{\text B}T/\epsilon_{AA}$
for different bonding energies. $x$ is the composition of species $1$. (a) $\epsilon_{BB}=\epsilon_{AA}$, $\epsilon_{AB}=1.05\epsilon_{AA}$, (b) $\epsilon_{BB}=\epsilon_{AA}$, $\epsilon_{AB}=0.95\epsilon_{AA}$, 
and (c) $\epsilon_{BB}=0.8\epsilon_{AA}$, $\epsilon_{AB}=0.85\epsilon_{AA}$. Left column: results from the cluster size distribution obtained using Monte Carlo simulations. Right column: results from the analysis of the percolation matrix with probabilities extracted from the Monte Carlo data (symbols)
and computed using Wertheim's theory (lines). The mixture percolates at temperatures below the green line. 
Species 1 (2) has percolated if $T$ is lower than the red (blue) line.}
\label{fig6}
\end{figure}

The full percolation behaviour is summarized in the composition-temperature percolation diagrams depicted in Fig.~\ref{fig6}. The percolation diagrams have been obtained via Monte Carlo simulations analyzing
the cluster size distribution (left) and theoretically, assuming that the network is a tree-like cluster and using the bonding probabilities measured in the simulations (symbols) and calculated using Wertheim's theory (lines). We first describe the global common features and then compare the two approaches. 

{\bf Symmetric mixture with $\mathbf{\epsilon_{AB}>\epsilon_{AA}}$}. In Fig.~\ref{fig6}a
we show the results for a symmetric mixture with $\epsilon_{BB}=\epsilon_{AA}$ and
$\epsilon_{AB}=1.05\epsilon_{AA}$. Three percolated states are possible below the
percolation threshold: two standard gels and a mixed gel. The mixed gel is stable
in a wide range of compositions, and it is the first type of gel stabilized
upon cooling the system. This is the expected behaviour as the whole mixture
will always percolate before the individual species percolate.
The temperature at the percolation threshold has a maximum for an equimolar mixture, $x=0.5$, which 
is the composition where the number of possible interspecies bonds, with the highest bonding
energy, is largest. At compositions $x\lesssim0.3$ ($x\gtrsim 0.7$) we
also find a gel of species $2$ ($1$) at sufficiently low temperature. 
Intraspecies bonds stabilize a standard gel, and they increase as the composition approaches that of a single component system.
Therefore, the temperature where the standard gel is stabilized increases as we approach the limit of pure systems $x\rightarrow0$ and
$x\rightarrow1$. In other words, the range in compositions where the mixed gel is stable decreases with the temperature (in the
range of temperatures analyzed here).

{\bf Symmetric mixture with $\mathbf{\epsilon_{AB}<\epsilon_{AA}}$}. If the intraspecies bonds have the highest energy there is a competition
between the entropy of mixing (which favours a mixed gel) and the internal energy of bonding (which favours a bicontinuous gel). An example is shown
in Fig.~\ref{fig6}b. At low temperatures the internal energy dominates and we observe a bigel (BG) in a range of compositions around $x=0.5$. 
The lower the temperature the wider the composition range where a BG is stable.
At high temperatures the entropy of mixing overcomes the internal energy of bonding and a mixed gel is stabilized. By contrast to the case with ${\epsilon_{AB}>\epsilon_{AA}}$,
the percolation temperature of the mixture exhibits a minimum at the equimolar composition $x=0.5$ rather than a maximum. This is simply because the number of interspecies
bonds is maximal at $x=0.5$, which are now the bonds with the lowest energy, and thus a lower temperature is required to  stabilize the bonds.
Standard gels are also present in this mixture near the composition of single component systems.

{\bf Asymmetric mixture}. The diagrams discussed above are not present only in the very special case of symmetric mixtures. An example is shown 
in Fig.~\ref{fig6}c where we consider an asymmetric mixture with bonding energies $\epsilon_{BB}=0.80\epsilon_{AA}$ and $\epsilon_{AB}=0.85\epsilon_{AA}$.
The diagram is now asymmetric with respect to the equimolar composition. Gel $1$ dominates over a gel of species 2 since  the $AA$ bonds are stronger than $BB$ bonds. 
Although $\epsilon_{AB}>\epsilon_{BB}$ there is a bigel at low temperatures. The reason is that the bonding energies satisfy $\epsilon_{AA}+\epsilon_{BB}<2\epsilon_{AB}$, i.e., 
it is energetically unfavorable to form two $AB$ bonds rather than one $AA$ and one $BB$. 
Therefore a bicontinuous gel should be stable at sufficiently low temperature. 

We next compare the percolation diagrams obtained directly using the cluster size distribution (left column of Fig.~\ref{fig6}) and using the percolation matrix and the bonding probabilities (right column of Fig.~\ref{fig6}). The derivation of the percolation matrix neglects the correlation between bonds and in particular neglects closed loops. The clusters are tree-like and their size is generally larger than that obtained from the simulations, where bond correlations and closed loops are present.
Consequently, the threshold temperatures predicted by the percolation matrix are higher than those obtained from the
simulations. This is, however, the only significant difference. In all cases, the results are qualitatively the same. We note that the bonding probabilities were obtained in two different ways: (i) theoretically using Wertheim's theory (solid lines in Fig.~\ref{fig6}) and (ii) directly from the simulations (symbols in Fig.~\ref{fig6}) by measuring the number of bonds of each type.
The resulting percolation diagrams are in perfect agreement and therefore we are confident that closed loops are the only relevant missing ingredient in our theoretical description.

We emphasize that we have analyzed the percolation diagram only. Some of the states considered above may be metastable against phase separation. For example, for the mixture with 
$\epsilon_{AB}<\epsilon_{AA}$ we observed signs of demixing at very low temperatures. We will return to this point in the Discussion. 

\subsection{Ternary mixtures}

In this section, we consider ternary mixtures by adding a third component (species $3$) with interaction sites of type $C$. 
The addition of a new species increases significantly the number of percolated states. In Fig.~\ref{fig7} we depict all gels labelled according to their strength against breaking bonds: 

\begin{figure}
\includegraphics[width=0.7\columnwidth]{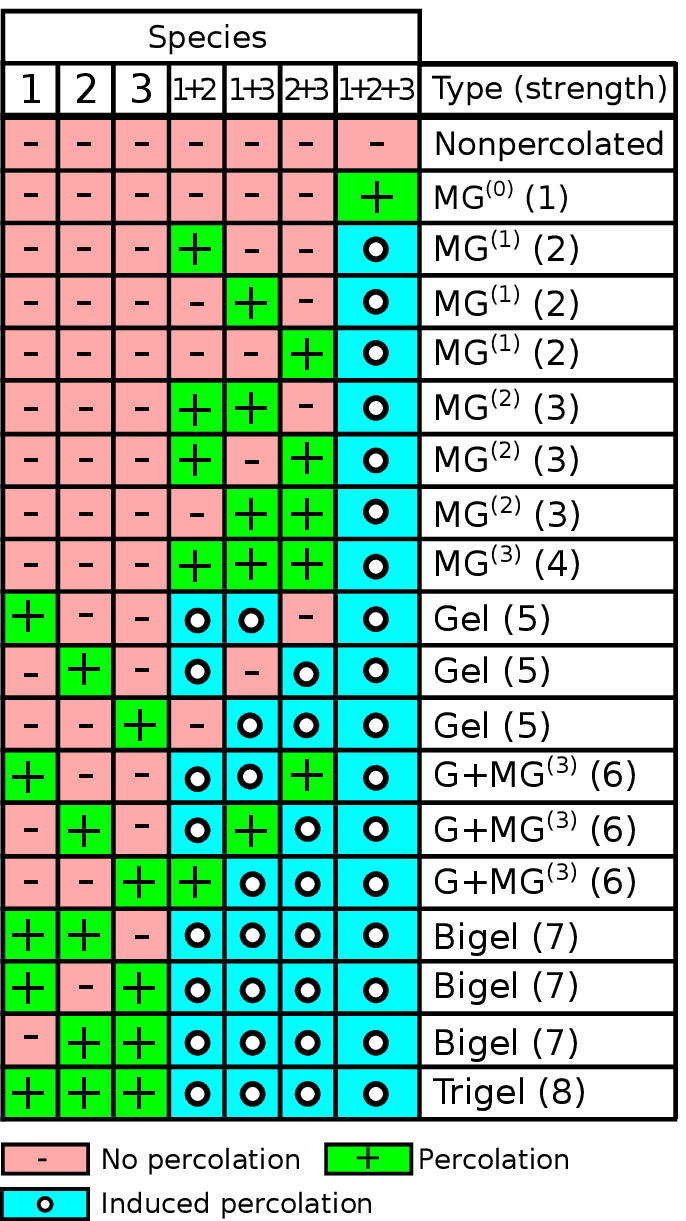}
\caption{Gels of a ternary mixture of patchy particles. The
number between brackets indicates the strength of the gel (the higher the number the stronger the gel).}
\label{fig7}
\end{figure}

{\noindent {\bf Mixed gels.}} The weakest type of percolated state is again a mixed gel. In a ternary mixture there are
four different mixed gels $\text{MG}^{(k)}$ with $k=0,1,2,3$ indicating the strength (the lower the index the weaker the gel).
In mixed gels the single species do not percolate. The difference between these gels relates to different percolating states of pairs of species. The weakest mixed gel is $\text{MG}^{(0)}$ where only the ternary mixture percolates. A subset of species, e.g., a mixture of species $1$ and $2$ does not percolate. Next, we find gels $\text{MG}^{(1)}$ where there is
one and only one pair of species that percolates (e.g., species $1$ and $2$, which in the following we denote by $\{1+2\}$),
and thus the ternary mixture also percolates.
There are three $\text{MG}^{(1)}$ gels related to the pair of species that percolates: $\{1+2\}$, $\{1+3\}$, and $\{2+3\}$. The bonds of the species that is not involved in percolation may be removed without breaking the global connectivity of the system.

Then we find $\text{MG}^{(2)}$ gels where two pairs of species percolate. Again, there are three different gels (see Fig.~\ref{fig7}). We can remove only one species without breaking the global connectivity, as in $\text{MG}^{(1)}$, but now we have two choices. For example, if the pairs $\{1+2\}$ and $\{1+3\}$ are percolated, we can break the bonds of either species $2$ or $3$ without affecting the global connectivity.

The strongest mixed gel is $\text{MG}^{(3)}$ where any pair of species has percolated. As in the previous cases we can break the bonds of only one species within the percolated state, but we can choose any of the three species.

{\noindent {\bf Standard gels.}} Here, one and only one of the species has percolated. Any pair of species where one component is the percolated species is
also percolated, and the ternary mixture has percolated as well. The pair of species that does not involve the percolated component has not percolated.
In a standard gel we can break the bonds of two species (the non percolated components) without affecting the connectivity
of the remaining particles.

{\noindent {\bf Standard plus mixed gels.}} In this case the $i$-th species ($i=1,2$ or $3$) has percolated and the pair of species $\{j+k\}$ with $j,k\neq i$ has also percolated.
As for a standard gel the connectivity is not lost if we break the bonds of species $j$ and $k$. In addition, we have the possibility of breaking the bonds of species $i$. 

{\noindent {\bf Bicontinuous gels.}} As for binary mixtures bicontinuous gels may also occur. Two species $i$ and $j$ percolate and any pair of species also percolates. Global connectivity is not affected by breaking the bonds of two species provided that one is species $k\neq i,j$.

{\noindent {\bf Trigel.}} The trigel is the strongest gel. The three species percolate and all pairs of species also percolate. The bonds of any two species may be broken without affecting the global connectivity of the system.

\begin{figure*}
\includegraphics[width=0.9\textwidth]{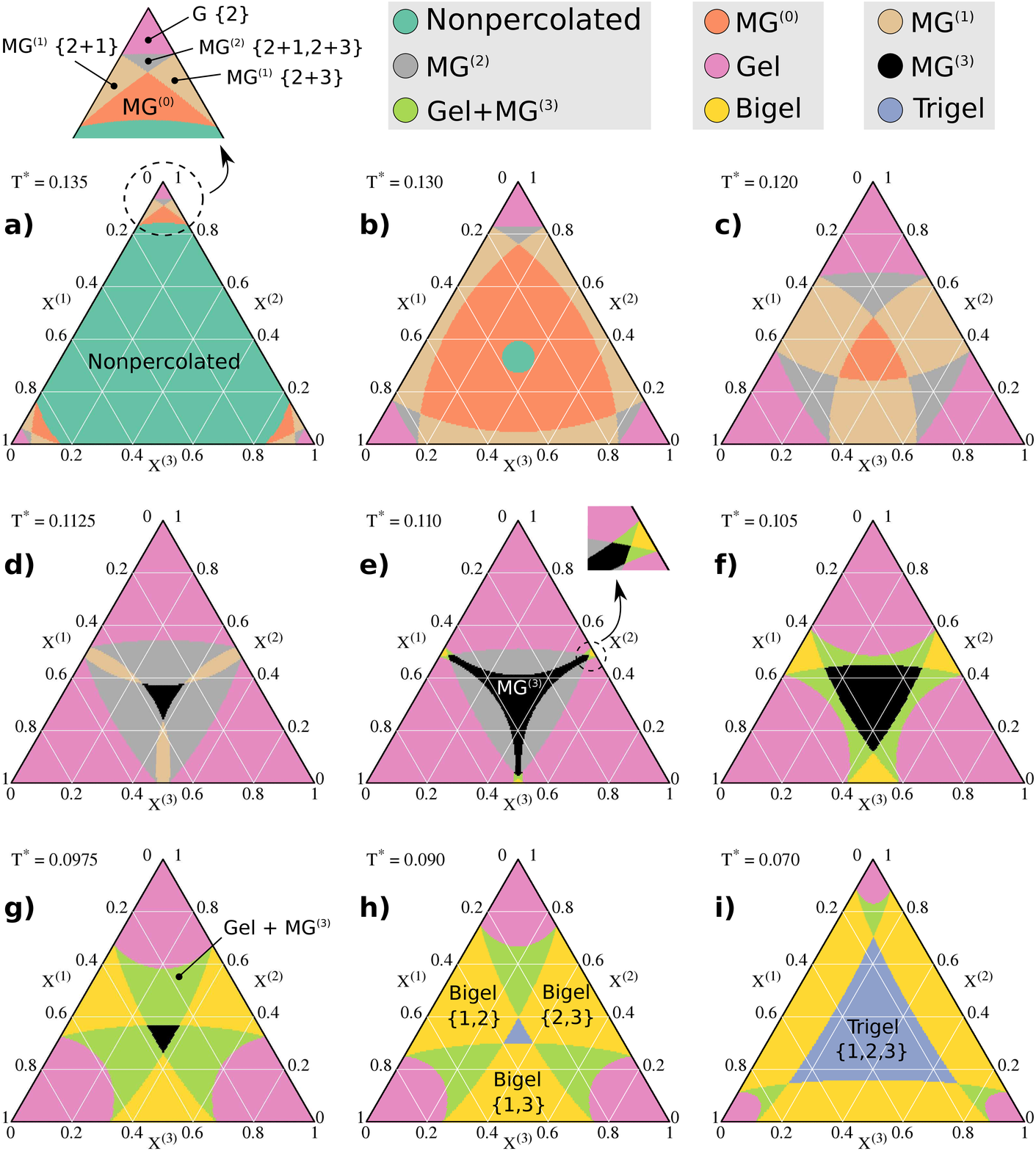}
\caption{Percolation diagrams (barycentric plots) of a ternary system of patchy colloids at different temperatures. In 
all cases the total packing fraction is $\eta=0.30$. Species 1,2 and 3 have three patches of
type A,B and C, respectively. Intraspecies bonds $\epsilon_{\alpha\alpha}=1$, $\alpha={\text{A},\text{B},\text{C}}$ are stronger
than the interspecies bonds $\epsilon_{\alpha\beta}=0.90\epsilon_{\alpha\alpha}$ with $\alpha\neq\beta$. The scaled-temperature is $T^*=k_{\text B}T/\epsilon_{\alpha\alpha}$. Each color
represents a different percolation state or gel (see the legend on the top of the figure). In panels (a) and (e) we also show close up views of selected regions. The
numbers between brackets indicate the species that have percolated (e.g., $\{2+1\}$ means that a mixture of species $2$ and $1$ percolates).
}
\label{fig8}
\end{figure*}

In binary mixtures when the intraspecies bonds are stronger than interspecies ones the percolation diagram is very rich and exhibits 
all types of gels. The reason is a competition between entropy and internal energy. The entropy of mixing promotes interspecies bonds which stabilize the mixed gel.
By contrast, the internal energy of bonding favour intraspecies bonds which are dominant in the bicontinuous gel. At high (low) temperature the entropy (internal energy) dominates
and a mixed (bicontinuous) gel is formed. The same competition occurs in ternary mixtures. The richest percolation behaviour also arises when interspecies bonds are weaker than
intraspecies ones. An example is shown in Fig.~\ref{fig8} where we plot the full percolation diagrams at different temperatures for a system with bonding  energies $\epsilon_{\alpha\alpha}=1$ (intraspecies) and 
$\epsilon_{\alpha\beta}=0.9\epsilon_{\alpha\alpha}$ (interspecies) with $\alpha,\beta=1,2,3$ and $\alpha\neq\beta$. Given the  excellent agreement between theory and simulations for binary
mixtures we have studied the percolation diagrams theoretically, and simulated only selected systems for each type of gel. 

Decreasing the temperature leads to an increase in the number of bonds, strengthening the percolated phases. As we will see, decreasing the temperature decreases the regions of the percolation diagram occupied by the weakest gels and increases the regions occupied by the strongest ones. Occasionally, two or more regions merge giving rise to a new stronger gel. This observation describes the most relevant features of the percolation behaviour of this type of ternary mixtures.

The highest temperature considered is $k_{\text B}T/\epsilon_{\text{AA}}=0.135$, see Fig.~\ref{fig8}a.
Although this is slightly below the percolation temperature of a single component system,
the percolation diagram is dominated by a non percolated state. 
Only at composition close to the pure systems does the ternary mixture percolate. In these
regions the percolation behaviour is very rich (see the close up view in Fig.~\ref{fig8}a). In the limit
$x_i\rightarrow1$ a gel of species $i$ occurs. By moving away from the pure species we find
the sequence of gels $\text{MG}^{(2)}-\text{MG}^{(1)}-\text{MG}^{(0)}$ and finally a non percolated state. To understand this
sequence consider the following. At $x_i=1$ the pure species has percolated and thus also the subsets $\{i+j\}$ and $\{i+k\}$. As we decrease $x_i<1$ we replace particles of species $i$ by particles of species $j$ and $k$ and therefore
reduce the number of intraspecies bonds $i$. Since the temperature is high (slightly below the percolation temperature), the 
decrease of intraspecies bonds $i$ disrupts the global connectivity of this species. Thus, the next gel state is
either a $\text{MG}^{(2)}$ where the subsets $\{i+j\}$ and $\{i+k\}$ percolate, or a $\text{MG}^{(1)}$ where only one of the previous
two subsets percolates. The compositions $x_j$ and $x_k$ determine which of the gels, $\text{MG}^{(2)}$ or $\text{MG}^{(1)}$, is stable.
Further decreasing the number of intraspecies bonds $i$ (i.e. moving away from the vertices of the percolation diagram) the subsets $\{i+j\}$ or $\{i+k\}$  become non percolating and the state changes to a $\text{MG}^{(0)}$ where only the ternary mixture percolates $\{1+2+3\}$. Finally, we
find a transition between a $\text{MG}^{(0)}$ and a non percolated state at compositions sufficiently far from the pure system. The reason is that approaching the center of the percolation diagrams increases the number of interspecies
bonds, which have the lowest energy, effectively decreasing the percolation temperature.

The range of compositions where the mixture percolates increases by further decreasing the temperature. At $k_{\text B}T/\epsilon_{\text{AA}}=0.130$ see Fig.~\ref{fig8}b, only
a small region close to the equimolar composition remains non percolated. There are no new percolated states with the previous ones occupying a larger 
area of the percolation diagram, especially the $\text{MG}^{(0)}$ gel that has grown at the expense of the non percolated state. 

At $k_{\text B}T/\epsilon_{\text{AA}}=0.120$, see Fig.~\ref{fig8}c, the mixture has percolated over the whole range of composition. The standard gels, $\text{MG}^{(2)}$, and $\text{MG}^{(1)}$ continued to grow
(occupying larger regions of the percolation diagram) at the expense of the $\text{MG}^{(0)}$ gel, which is the weakest and has been reduced to a small region around the equimolar composition. 

At $k_{\text B}T/\epsilon_{\text{AA}}=0.1125$, see Fig.~\ref{fig8}d, the $\text{MG}^{(0)}$ gel has disappeared. The three regions where the $\text{MG}^{(2)}$ gel is stable have merged in the center of
the percolation diagram, giving rise to a new state, the $\text{MG}^{(3)}$ gel, where any pair of species percolates. The standard gels, which are the strongest
at this temperature, continue to grow at the expense of the weakest gel, $\text{MG}^{(1)}$.  

At the next temperature $k_{\text B}T/\epsilon_{\text{AA}}=0.11$
(see Fig.~\ref{fig8}e), the $\text{MG}^{(1)}$ has disappeared and new states occur. Two standard gels merge at compositions close to $x_i\approx x_j \approx 0.5$ giving 
rise to bicontinuous gels of species $i$ and $j$. Next to the regions of stability of the bigels, a standard gel and a $\text{MG}^{(3)}$ gel have merged and a Gel+$\text{MG}^{(3)}$
phase has appeared. The weakest gel is now $\text{MG}^{(2)}$ and the strongest is no longer a standard gel but a bigel.

Bigels and Gel+$\text{MG}^{(3)}$ states grow by further decreasing $T$, see Fig.~\ref{fig8}f, and completely replace the weak $\text{MG}^{(2)}$ gel state. At this point, the central $\text{MG}^{(3)}$
gel is the weakest, and starts to shrink. At $k_{\text B}T/\epsilon_{\text{AA}}=0.0975$, see Fig.~\ref{fig8}g, the $\text{MG}^{(3)}$ is confined to a small area near the equimolar
composition. The standard gels also shrink as the stronger gels grow.

The trigel, the strongest of all gels, appears at the next temperature $k_{\text B}T/\epsilon_{\text{AA}}=0.112$ as a result of the merging of the three bigels close to the equimolar composition, 
see Fig.~\ref{fig8}h. If we continue to cool the system the trigel grows at the expense of the weaker gels, see Fig.~\ref{fig8}i. No further topological change of the percolation diagram occurs at lower temperatures.

\begin{figure*}
\includegraphics[width=0.95\textwidth]{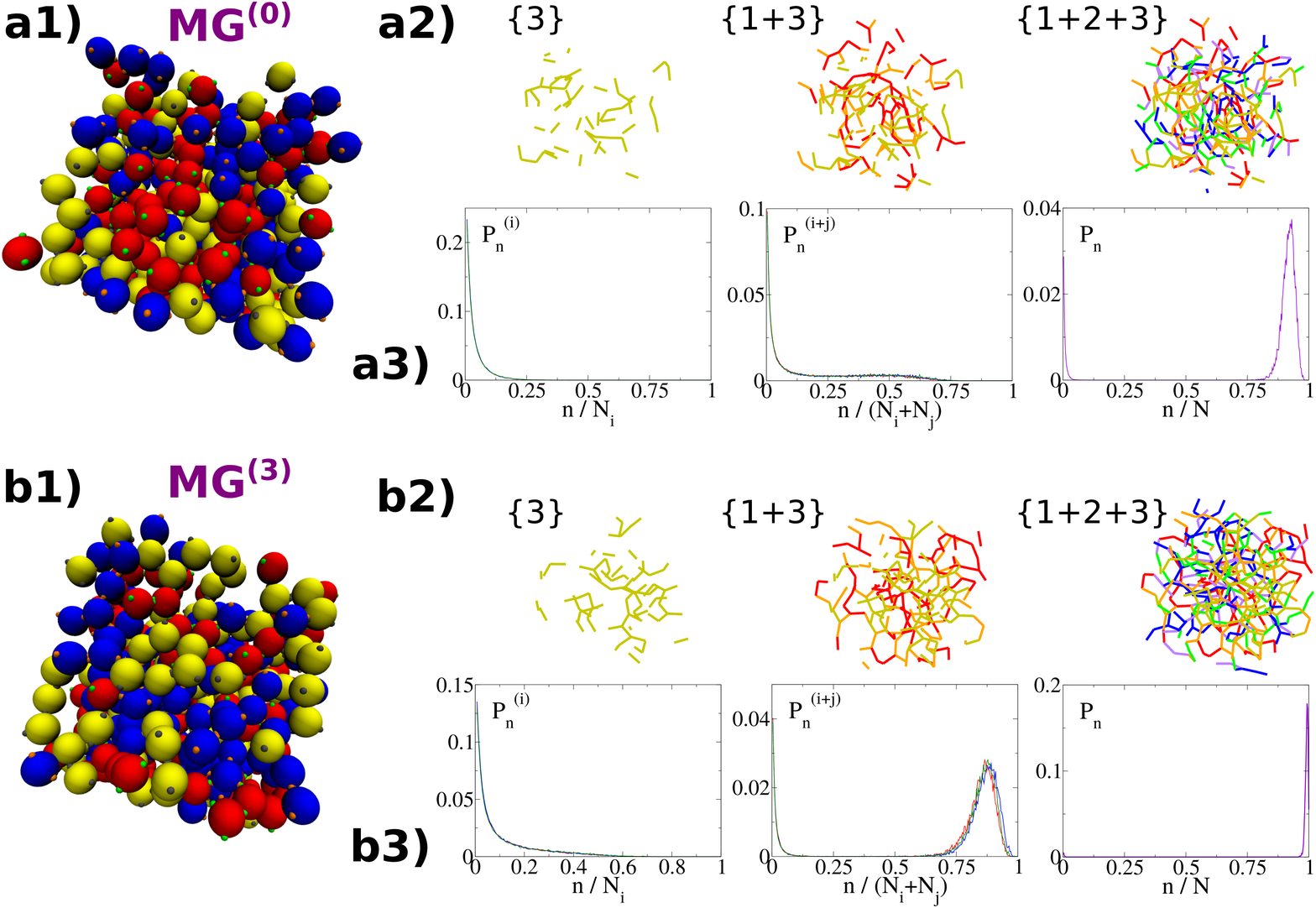}
\caption{Simulation results of a ternary mixture exhibiting a $\text{MG}^{(0)}$ gel (top panels) at $k_{\text B}T/\epsilon_{\text{AA}}=0.11$ and a $\text{MG}^{(3)}$ gel
(bottom panels) at $k_{\text B}T/\epsilon_{\text{AA}}=0.095$. In both cases the ternary mixture is equimolar ($x_i=1/3$ for $i=1,2$, and $3$) and the packing fraction
is $\eta=0.30$. The simulation box is cubic with length $9\sigma$. The intraspecies bonding energies are $\epsilon_{\alpha\alpha}=1$, $\alpha={\text{A},\text{B},\text{C}}$, stronger than interspecies bonds, $\epsilon_{\alpha\beta}=0.90\epsilon_{\alpha\alpha}$ with $\alpha\neq\beta$.
(a1) and (b1): Typical simulation snapshots. (a2) and (b2): Graphical representation of the bonds; each
line connects the center of mass of two bonded particles of species $3$ (left), of species 1 and 3 
(middle), and of any of the species (right). (a3) and (b3): Probability of finding a cluster of size $n$
of particles of species $i$ (left), of species $i$ or $j$ (middle), and of any species (right) as 
a function of the cluster size (normalized by the corresponding number of particles).
}
\label{fig9}
\end{figure*}

In addition to the theoretical percolation diagrams, we have simulated selected systems at different compositions and temperatures. In all cases we have found perfect agreement of the sequences of percolated states as the temperature decreases. For example, by cooling an equimolar system ($x_1=x_2=x_3=1/3$) the simulations exhibit the sequence non percolated-$\text{MG}^{(0)}$-
$\text{MG}^{(3)}$-Trigel, as predicted by the theory. As an example we show in Fig.~\ref{fig9} the simulation results for the gels $\text{MG}^{(0)}$ (top panels) and $\text{MG}^{(3)}$ (bottom panels), 
obtained by decreasing  the temperature from a non percolated state. The only relevant difference was the temperature at which each phase appears. We found, for example, that a trigel
is stable at temperatures $k_{\text B}T/\epsilon_{\text{AA}}\lesssim0.075$, whereas theoretically we observe stable trigels at $k_{\text B}T/\epsilon_{\text{AA}}\approx0.09$.
As for binary mixtures, the theory overestimates the percolation temperature of each type of gel. This discrepancy is attributed to the neglect of closed loops in the theoretical description. 

We have also computed the percolation diagrams for other sets of bonding energies, and the sequence of states and
topology of the percolation diagrams is always similar (the symmetry is obviously lost in cases with
asymmetric interaction energies). The most significant differences occur if all interspecies bonds
are energetically more favorable than the intraspecies bonds. In this case, the trigel and the bigels are not 
stable since they require a predominance of interspecies bonds. Also,   
the percolation temperature of the weakest $\text{MG}^{(0)}$ gel has a maximum at the equimolar composition, instead of a minimum
(like in Fig.~\ref{fig8}a-b). This is analogous to what happens in binary mixtures.

\section{Discussion}

We have considered the percolation transition in binary and ternary mixtures of patchy colloidal particles, at fixed packing fraction ($\eta=0.30$). The results of Wertheim's theory and Monte Carlo simulations indicate that a pure system of patchy particles with three identical patches at this packing fraction is thermodynamically stable over the whole range of temperature~\cite{:/content/aip/journal/jcp/128/14/10.1063/1.2888997}, i.e., there is no liquid-gas or
liquid-solid phase separation at $\eta=0.30$. The addition of other species might change the stability of the fluid phase. For
example, we expect demixing at sufficiently low temperatures if the intraspecies bonds are stronger than the interspecies
ones, in line with Wertheim's theory for binary mixtures~\cite{herasBC}. Nevertheless, in~\cite{herasBC} we have shown that it is possible to design more
complex patchy colloids (with different types of patches) in order to suppress the phase separation and stabilize the desired gels. In addition,
arrested phase separation might occur, and effectively stabilize the gel. An example is the formation of bigels in mixtures of DNA-coated colloids 
as a results of arrested demixing~\cite{Varrato20112012}.

Our theoretical description neglects the effect of closed loops. Nevertheless, simulations (closed loops included) and theory (no closed loops) are in semi-quantitative agreement. The effect of loops in the present model simply shifts the percolation threshold to lower temperatures. The bonding probabilities are almost unaffected by the presence of closed loops. This is not a general feature, as closed loops could have a major impact in the percolation behaviour of the system. This will occur in systems where the formation of ring-like structures is favoured. For example, in a system of particles with two patches of type A near the poles and several patches of type B on the equator~\cite{PhysRevLett.111.168302} the formation of
rings through AA bonds leads to a lower critical point in the liquid-vapor binodal. See~\cite{PhysRevE.87.052307,C5SM00559K} for recent discussions of
extensions of Wertheim's theory to include the effect of loops.

We have characterized the percolated states or gels according to the species that are percolated. For example, in a bigel there are two
species that percolate independently. One can also characterize the states according to the types of bonds that are percolated. In this
case one can for example distinguish between two types of bigels, one bigel where there are only two percolated networks formed
by intraspecies bonds and another bigel where the interspecies bonds also form a percolated network. 

Our model may help to understand network formation in other systems such as ABC block terpolymers~\cite{doi:10.1021/ma9009593},
and ternary polymer blends~\cite{doi:10.1021/ma500302k}. In these systems one can find percolated states analogous to those described here. For example,
the percolation behaviour of a ternary polymer blend made of high-density polyethylene, polystyrene, and PMMA is topologically similar
to the low temperature percolation behaviour of the ternary mixture analyzed here~\cite{Ravati20104547} (Fig.~\ref{fig8}h). 

Future studies will focus on the percolation behaviour under gravity. A gravitational field is, in general, unavoidable in 
sedimentation experiments, and might have a profound effect on colloidal mixtures~\cite{Sedi}.
For example, in a mixture of patchy colloids with two and three patches (all of the same type)
gravity induces the formation of complex sequences in sedimentation-diffusion-equilibrium such as the presence of re-entrant percolation~\cite{PhysRevE.93.030601}.

\begin{acknowledgments}
We thank Jose M. Tavares for useful discussions.
This work was support in part by the Portuguese 
Foundation for Science and Technology (FCT) through 
projects EXCL/FIS-NAN/0083/2012 and UID/FIS/00618/2013.
\end{acknowledgments}

\end{document}